\documentclass{camera}
\usepackage{graphicx}  
\usepackage{float}

\begin{document}

%
\title{Chiral nucleon-nucleon forces in nuclear structure calculations}

%
\author{L. Coraggio$^1$, A. Gargano$^1$, J. W. Holt$^2$, N. Itaco$^1$,
  R. Machleidt$^3$, L. E. Marcucci$^4$, \and F. Sammarruca$^3$}
\organization{$^1$ Istituto Nazionale di Fisica Nucleare, Via Cintia - I-80126 Napoli, Italy\\
$^2$ Physics Department, University of Washington, Seattle, WA 98195, USA \\
$^3$ Department of Physics, University of Idaho, Moscow, ID 83844, USA\\
$^4$ Dipartimento di Fisica ``Enrico Fermi'', Universit\`a di Pisa,
and Istituto Nazionale di Fisica Nucleare, Largo B. Pontecorvo 3 -
I-56127 Pisa, Italy}

\maketitle

\begin{abstract}
Realistic nuclear potentials, derived within chiral perturbation
theory, are a major breakthrough in modern nuclear structure theory,
since they provide a direct link between nuclear physics and its
underlying theory, namely the QCD.
As a matter of fact, chiral potentials are tailored on the low-energy
regime of nuclear structure physics, and chiral perturbation theory
provides on the same footing two-nucleon forces as well as many-body
ones.
This feature fits well with modern advances in ab-initio methods and
realistic shell-model.
Here, we will review recent nuclear structure calculations, based on
realistic chiral potentials, for both finite nuclei and infinite
nuclear matter.
\end{abstract}

%

\section{Introduction} \label{intro}
The last decade has seen chiral EFT emerging as a powerful tool to
describe the interaction among hadrons in a  low-energy regime in a
systematic and model-independent way (see Refs.\ \cite{ME11,EHM09} for
recent reviews). 

Generally, in nuclear many-body systems the nuclear momenta are on the
order of the pion mass, so the construction of chiral nuclear
potentials is based on an expansion in powers of this soft scale ($Q
\sim m_\pi$) over the hard scale, which is set by the typical hadron
masses $\Lambda_\chi \sim m_\rho \sim 1$~GeV. 
The latter is also known as the chiral-symmetry breaking scale. 
The EFT must have a firm link with quantum chromodynamics (QCD) so to
emerge above the level of phenomenology.
This is obtained having the EFT undergo all relevant
symmetries of the underlying theory, in particular, the broken chiral
symmetry of low-energy QCD \cite{Wei79}.

Nuclear potentials based on chiral perturbation theory (ChPT)
\cite{ME11,EM03,EGM05} are at present largely employed so to bridge
the QCD, that is the fundamental theory of
strong interactions, to the physics of the atomic nucleus.
A relevant apect of ChPT is that nuclear two-body forces, many-body 
forces, and currents \cite{ME11,Wei92,Kol94} are generated on an equal
footing. 
As a matter of fact, some low-energy constants (LECs) which appear in
the two-nucleon-force (2NF), fitted to two-nucleon data, belong also
to the expressions of the three-nucleon forces (3NF), four-nucleon
forces (4NF), and electroweak currents.

In this contribution we will review about calculations for many-body
systems, both finite nuclei and infinite nuclear matter, performed
starting from nuclear forces derived within the ChPT.
More precisely, we consider a low-momentum chiral potential
constructed at next-to-next-to-next-to-leading order (N$^3$LO) using a
sharp cutoff $\Lambda = 414$ MeV \cite{Coraggio07}, dubbed as
N$^3$LOW.

The paper is organized as follows. 
In Sec. \ref{theory}, we briefly outline the theoretical framework
of our many-body calculations.
More precisely, we will present a few details of the chiral potential
we have employed, of the theory underlying the derivation of an
effective hamiltonian for a shell-model calculation, and of the
perturbative expansion of the energy per particle in infinite neutron-
and symmetric-nuclear matter. 

Sec. \ref{finitenuclei} is dedicated to the presentation of results of
realistic shell-model calculation for medium-mass nuclei outside
$^{14}$C and $^{16}$O closed-shell nuclei, starting from N$^3$LOW
chiral potential. 
The results of the calculation of the equation of state (EOS) for
neutron- and symmetric-infinite matter are reported in Sec. \ref{EOS},
where we also focus on the comparison of the calculated symmetry
energy with the empirical values.

Finally, in Sec. \ref{conclusions} our conclusions are reported.

\section{Outline of calculations} \label{theory}
First of all, we start this section with a brief description of the
chiral potential N$^3$LOW.
This potential has been introduced in Ref. \cite{Coraggio07}, so to be
compared directly with a $V_{\rm low-k}$ derived from the chiral
potential of Entem and Machleidt \cite{EM03}, renormalized within a
cutoff momentum $\Lambda=2.1$ fm$^{-1}$ \cite{Bogner02}.

In the $V_{\rm low-k}$ approach, the original potential $V_{NN}$ is
smoothed by integrating out the high-momentum modes of  $V_{NN}$ down
to a cutoff momentum $\Lambda$. 
This is achieved by a unitary transformation as the one, for example,
suggested by Lee and Suzuki \cite{Lee80,Bogner02}.

Basically, the $V_{\rm low-k}$ approach to nuclear structure physics
is successful because the dynamics which rules nuclear physics can be
described in the framework of a low-energy EFT. 

As already mentioned in Sec. \ref{intro}, this nuclear EFT is
characterized by the spontaneously broken chiral symmetry, and the
degrees of freedom which are relevant for nuclear physics, nucleons
and pions.

In order to reach a proper convergence rate, the ChPT expansion is
valid only for momenta $ Q < \Lambda_\chi \simeq 1$ GeV, where
$\Lambda_\chi$ denotes the chiral symmetry breaking scale.
Nucleon-nucleon ($NN$) potentials derived from ChPT are typically
multiplied by a (non-local) regulator function $f(p',p) = \exp
[-(p'/\Lambda)^{2n} - (p/\Lambda)^{2n}]$, where $\Lambda \simeq 0.5$
GeV is a typical choice for the cutoff scale.
In terms of a relative-momentum cutoff, this means that present chiral
$NN$ potentials \cite{EM03,EGM05} typically apply values for
$\Lambda$ around 2.5 fm$^{-1}$.

On these grounds, in Ref. \cite{Coraggio07} a chiral N$^3$LO $NN$
potential has been considered, using the regulator function in the
above expression with $n=10$ and $\Lambda = 414$ MeV, so to be
confronted directly with a $V_{\rm low-k}$ whose cutoff momentum is
$\Lambda=2.1$ fm$^{-1}$. We have dubbed this potential ${\rm
  N^3LOW}$.

One advantage of this potential, respect to a $V_{\rm low-k}$ is that
it is given in analytic form. 
The analytic expressions are the same as for the ``hard'' N$^3$LO
potential by Entem and Machleidt of 2003 \cite{EM03}. 

\begin{figure}[ht]
\begin{center}
\includegraphics[scale=0.45,angle=0]{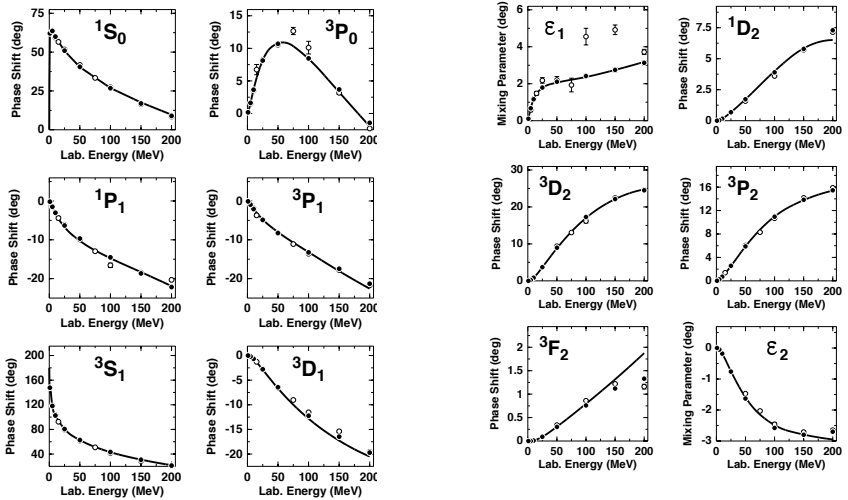}
\end{center}
\caption{Phase parameters of neutron-proton scattering
up to 200 MeV laboratory energy.
The solid lines show the predictions by the ${\rm N^3LOW}$ potential.  
The solid dots and open circles represent the Nijmegen multienergy
$np$ phase shift analysis and the GWU/VPI single-energy $np$ analysis
SM99, respectively.} 
\label{fig1} 
\end{figure}

The N$^3$LOW potential reproduces accurately the
empirical deuteron binding energy, the experimental low-energy
scattering parameters, and the empirical phase-shifts of $NN$
scattering up to at least 200 MeV laboratory  energy (see
fig. \ref{fig1}). 

Another important issue are many-body forces. 
One great advantage of ChPT is that it generates nuclear two- and
many-body forces on an equal footing.
Most interaction vertices that appear in the 3NF and in the
four-nucleon force (4NF) also occur in the two-nucleon force (2NF). 
The parameters carried by these vertices are fixed in the
construction of the chiral 2NF. 
Consistency requires that for the same vertices the same parameter
values are used in the 2NF, 3NF, 4NF, \ldots .

If one considers the perturbative expansion of the 3NF at N$^2$LO, then
only two extra-parameters need to be fixed, namely the low-energy
constants (LECs) dealing with the one-pion exchange three-nucleon force
$V_{3N}^{1\pi}$ and the contact three-nucleon force $V_{3N}^{\rm
  cont}$ at N$^2$LO.
These 3NF LECs have been adjusted to $A = 3$ observables, more
precisely so to reproduce the $^3$H and $^3$He binding energies
together with the triton half-life (specifically the Gamow-Teller
matrix element) \cite{Marcucci12,Coraggio14a}.

We have employed the N$^3$LOW potential in calculations for both
finite nuclei \cite{Coraggio07,Coraggio10,Coraggio14b} - within the
framework of the realistic shell model - and for infinite nuclear
matter \cite{Coraggio14a,Coraggio13}.

Let us start introducing the realistic shell-model.

We intend as a realistic shell-model, calculations for finite nuclei
where the effective shell-model hamiltonian, namely the
single-particle energies (s.p.e.) and two-body matrix elements (TBME)
of the residual interaction, has been derived from a realistic $NN$ force
\cite{Coraggio12}, such as the N$^3$LOW potential.

Starting from this potential, the shell-model effective hamiltonian is
derived within the many-body perturbation theory, as
developed by Kuo and coworkers through the 1970s
\cite{Kuo90}.
More precisely, we have used the well-known $\hat{Q}$-box plus
folded-diagram method \cite{Coraggio12,Kuo90}, where the $\hat{Q}$-box
is a collection of one- and two-body irreducible valence-linked Goldstone
diagrams.
In the following calculations the $\hat{Q}$-box includes  all diagrams
up to third order \cite{Coraggio12},  and the
folded-diagram series is summed up to all orders using the Lee-Suzuki
iteration method \cite{Lee80}.
We sum over a number of intermediate states between successive vertices whose
unperturbed excitation energy is less than a value $E_0=E_{max}$,
which is sufficiently large to ensure that both s.p.e. and TBME are
almost independent from the choice of $E_{max}$.

As regards the calculation of EOS of infinite nuclear matter, we
calculate the ground state energy (g.s.e.) per particle within the
framework of many-body perturbation theory.
More precisely, the g.s.e. is expressed as a sum of Goldstone diagrams
up to third order.

The effects of the 3NF are taken into account via a density-dependent
two-body potential ${\overline V}_{3N}$, that is added to the chiral
N$^3$LO potential $V_{2N}$.
This potential is obtained by integrating one nucleon up to the Fermi
momentum $k_F$, thus leading to a density-dependent two-nucleon
interaction ${\overline V}_{3N}(k_F)$.
Presently, the analytic expressions for ${\overline V}_{3N}$
\cite{HKW10} has been derived only for the N$^2$LO 3NF, which is the
one we take into account in the calculation of our EOS.
In order to consider the correct combinatorial factors of the
normal-ordering at the two-body level of the 3NF, the matrix elements
of ${\overline V}_{3N}(k_F)$ have to be multiplied by a factor 1/3 in
the first-order Hartree-Fock (HF) diagram, and by a factor 1/2  in the
calculation of the HF single-particle energies \cite{Coraggio13,Coraggio14a}.

The analytic expressions of first-, second-, and third-order
particle-particle ($pp$) and hole-hole ($hh$) contributions, together
with the one of single-particle HF potential, are  reported in
Ref. \cite{Coraggio13}.
The implicit expression of the third-order particle-hole $ph$ diagram
- also known as the ring diagram - can be found in
Ref. \cite{MacKenzie69}.

\section{Shell-model results}\label{finitenuclei}
In Ref. \cite{Coraggio10}, we have presented a shell-model description of heavy
carbon isotopes, using a fully microscopic approach, starting from
N$^3$LOW potential and using $^{14}$C as a closed core.
As mentioned in sec. \ref{theory}, both the single-particle energies and residual
two-body interaction have been derived within the many-body
perturbation theory.
Our calculations have led to a successful description of these isotopes
when approaching the neutron dripline. 
More precisely, our results have reproduced the disappearance of the $N=14$
subshell closure that is present in the oxygen chain, and predicted the
$N=16$ one.

\begin{figure}[ht]
\begin{center}
\includegraphics[scale=0.45,angle=0]{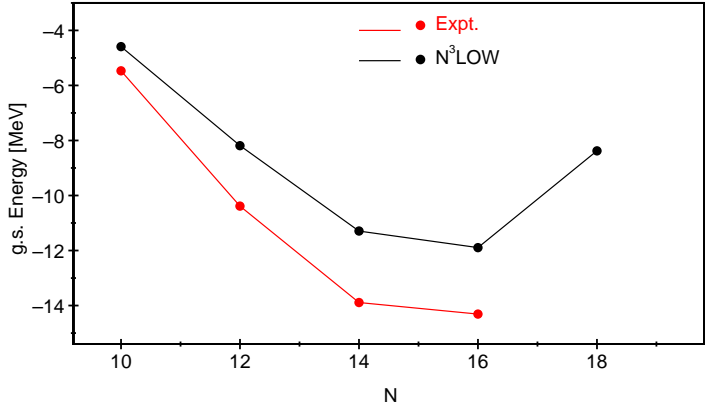}
\end{center}
\caption{(Color online)  Experimental and calculated ground-state
  energies for carbon isotopes from $A=16$ to 24. $N$ is the number of
  neutrons. See text for details.}
\label{fig2} 
\end{figure}

Moreover, the shell-model calculations have predicted that $^{21}$C is
unstable against one-neutron decay, fitting the picture of $^{22}$C as
a Borromean nucleus.
It is worth mentioning that the latter is one of the most exotic
nuclei of the 3000 known isotopes, its $N/Z$ ratio being 2.67,  and
can be considered the heaviest Borromean nucleus ever observed.
We reproduce successfully the fact that $^{22}$C is the
last bound isotope (see fig. \ref{fig2}).

\begin{figure}[hb]
\begin{center}
\includegraphics[scale=0.35,angle=0]{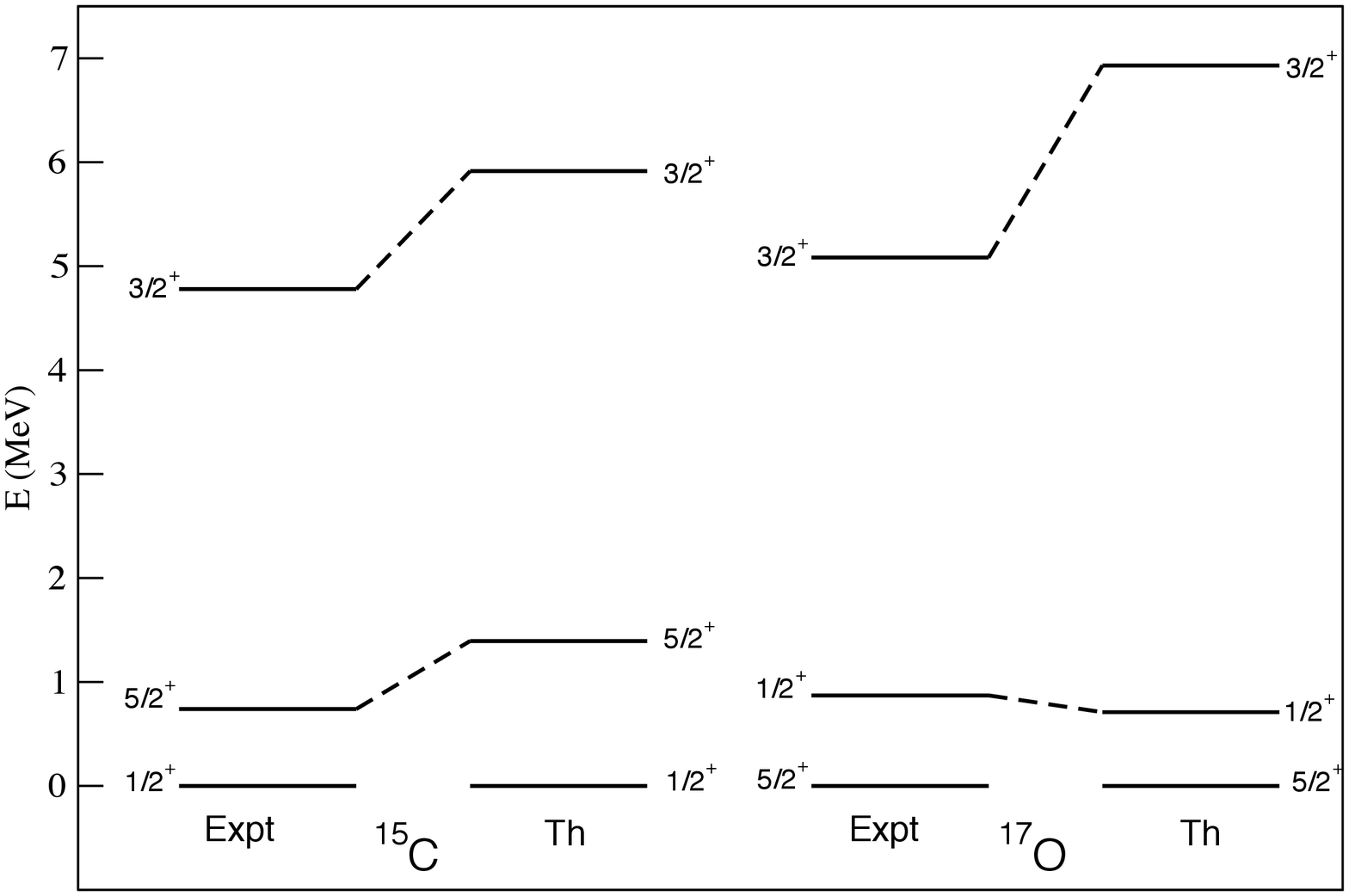}
\end{center}
\caption{Experimental \cite{ensdf} and theoretical single-particle
  states of $^{15}$C and $^{17}$O.}
\label{fig3} 
\end{figure}

 In ref. \cite{Coraggio10} we investigated only the identical-particle
channel of the effective hamiltonian. 
Here, we point out that also the proton-neutron channel features well,
as testified in fig. \ref{fig3} where the experimental and
theoretical single-particle spectra of $^{15}$C and $^{17}$O are
reported.
\begin{figure}[H]
\begin{minipage}{14pc}
\begin{center}
\includegraphics[scale=0.34,angle=0]{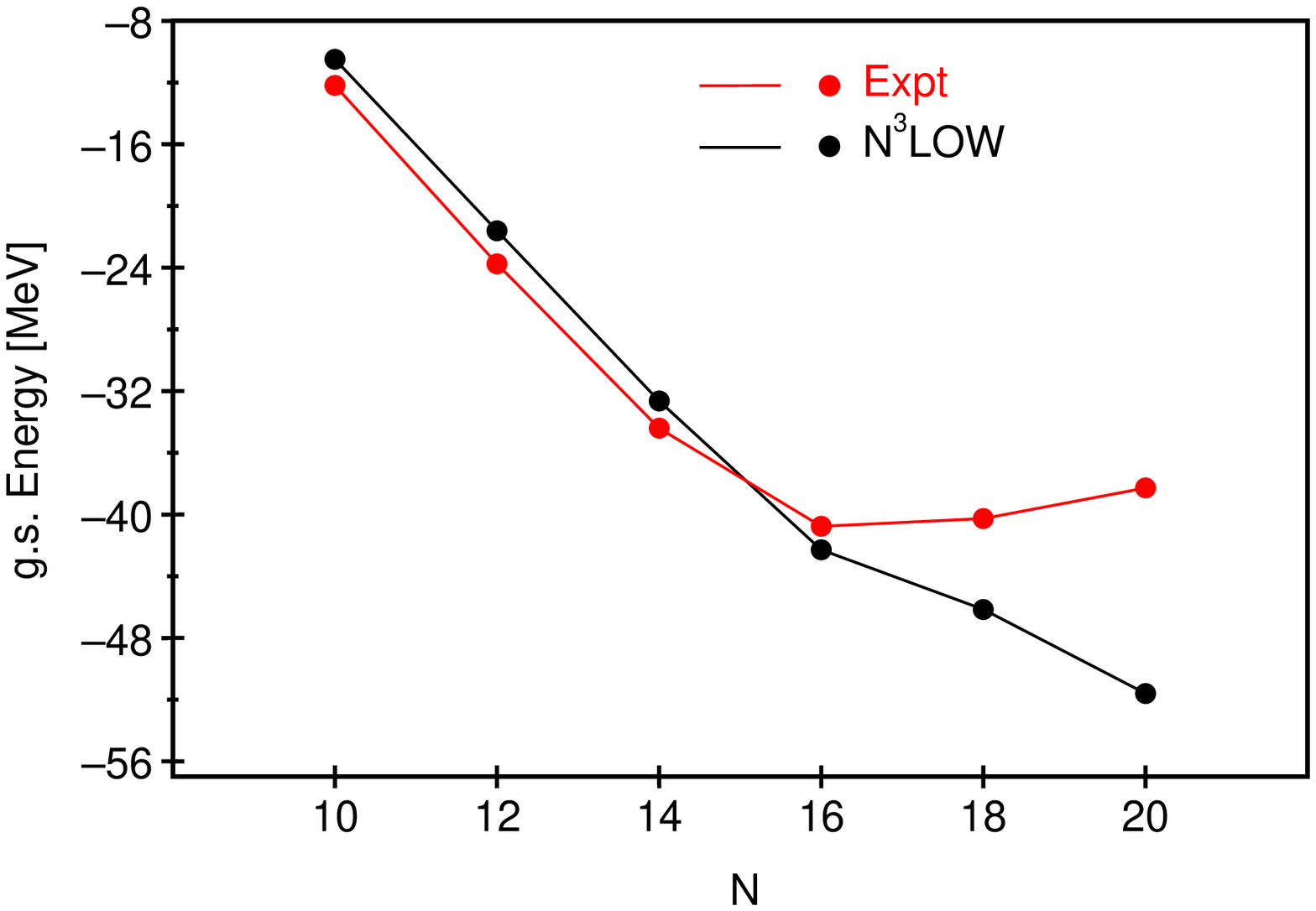}
\caption{(Color online)  Experimental and calculated ground-state
  energies for oxygen isotopes from $A=18$ to 24. $N$ is the number of
  neutrons.}
\label{fig4}
\end{center}
\end{minipage}\hspace{2pc}%
\begin{minipage}{14pc}
\begin{center}
\hspace{-1pc}
\includegraphics[scale=0.30,angle=90]{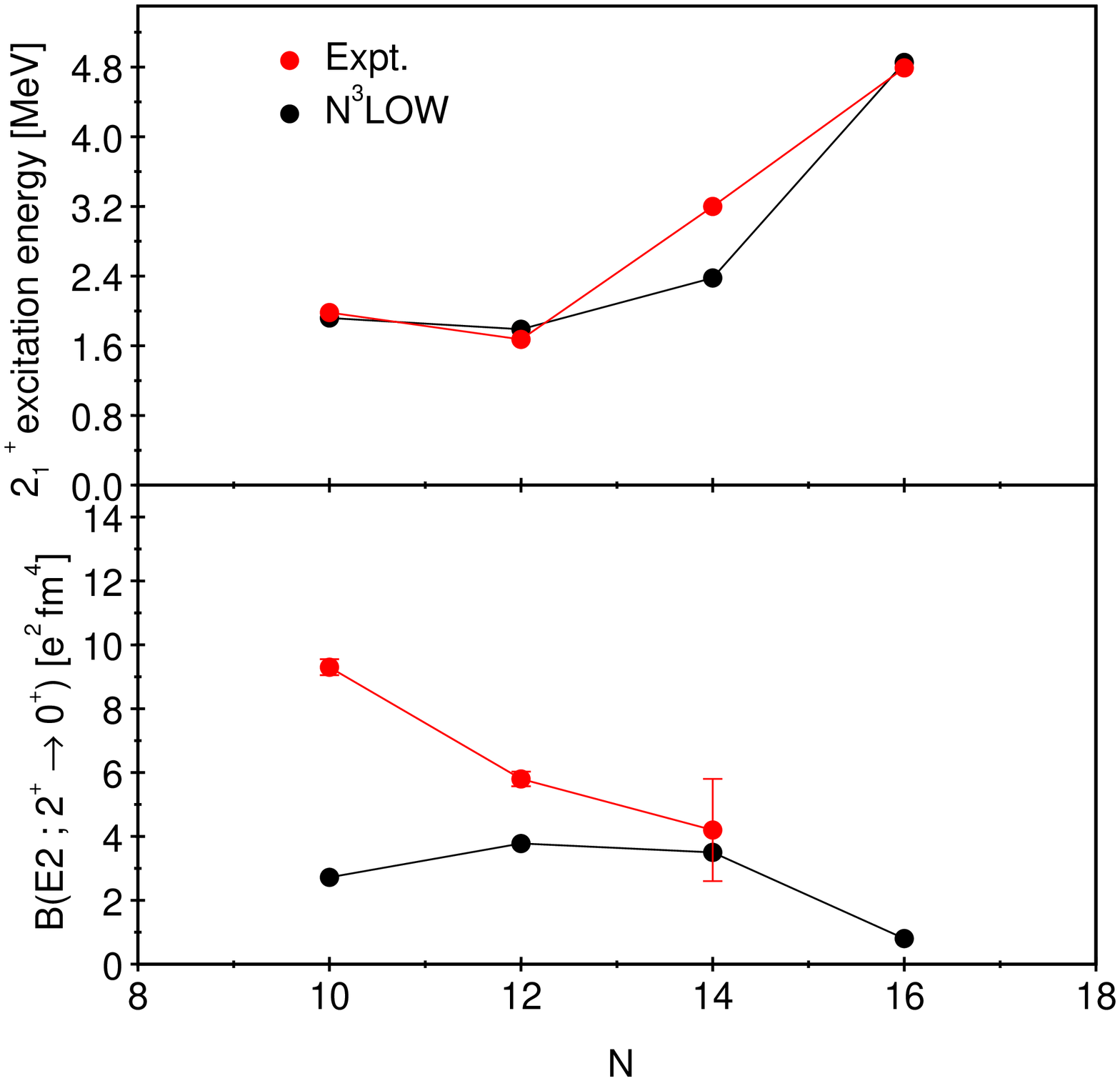} 
\caption{(Color online) Experimental \cite{ensdf} and
  calculated excitation energies of the yrast $J^{\pi}=2^+$ states and
  $B(E2;2^+_1 \rightarrow 0^+_1)$  transition rates for oxygen
  isotopes.}
\label{fig5} 
\end{center}
\end{minipage} 
\end{figure}

As can be seen the inversion of the $J^{\pi}=1/2^+$ and $5/2^+$ states
is correctly reproduced, thanks to the proton-neutron monopole
component of the effective hamiltonian.

We have employed the N$^{3}$LOW potential also for calculations of
nuclei belonging to the $sd$-shell, using $^{16}$O as a closed core.
In ref. \cite{Coraggio14c} we reported only results for
two-valence-particle nuclei $^{18}$O and $^{18}$F.
Here we report in fig. \ref{fig4} the calculated ground-state
energies, with respect to $16$O, of even-mass oxygen isotopes and
compare them with the experimental ones.
As can be observed, the agreement is quite remarkable up to $^{24}$O,
then the theory predicts a bound $^{26}$O, at variance with experiment.
This probably traces back to the fact that we have not included the
effect of three-body forces in the derivation of our effective
hamiltonian \cite{Otsuka10a}

In fig. \ref{fig5} we have also reported the experimental and
calculated excitation energies of the yrast $J^{\pi}=2^+$ states and
$B(E2;2^+_1 \rightarrow 0^+_1)$  transition rates.

\section{Infinite nuclear matter}\label{EOS}
As mentioned previously, we calculate the energy per particle of
infinite nuclear matter in the framework of the Goldstone expansion
taking into account the effects of the N$^2$LO 3NF  via the
introduction of a density-dependent two-body potential  ${\overline
  V}_{NNN}(k_F)$. 

\begin{figure}[hb]
\begin{minipage}{14pc}
\begin{center}
\includegraphics[scale=0.27,angle=0]{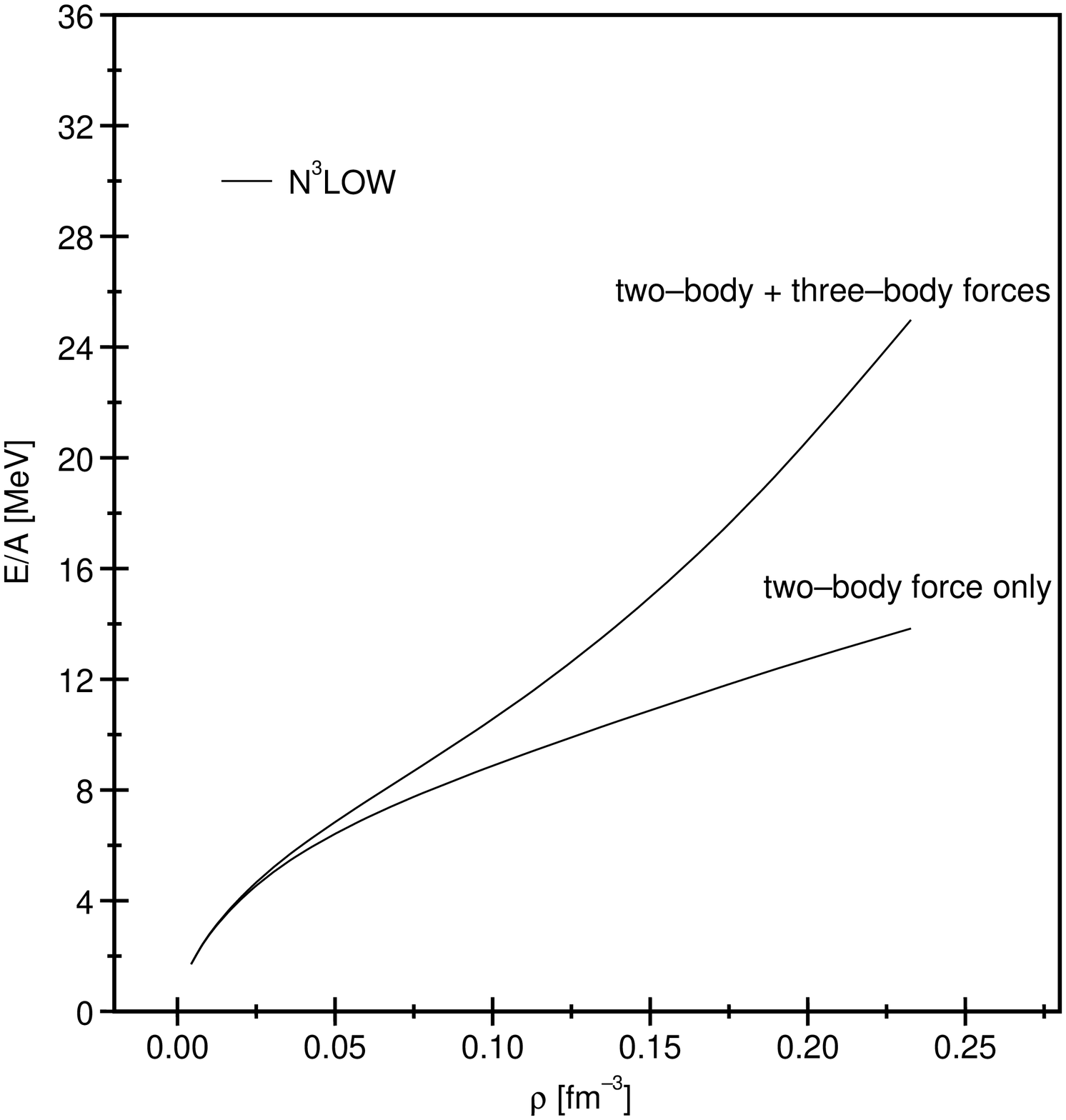}
\caption{\label{Fig6} Results obtained for the PNM energy per 
nucleon, with and without 3NF effects.}
\end{center}
\end{minipage}\hspace{2pc}%
\begin{minipage}{14pc}
\begin{center}
\hspace{-1pc}
\includegraphics[scale=0.27,angle=0]{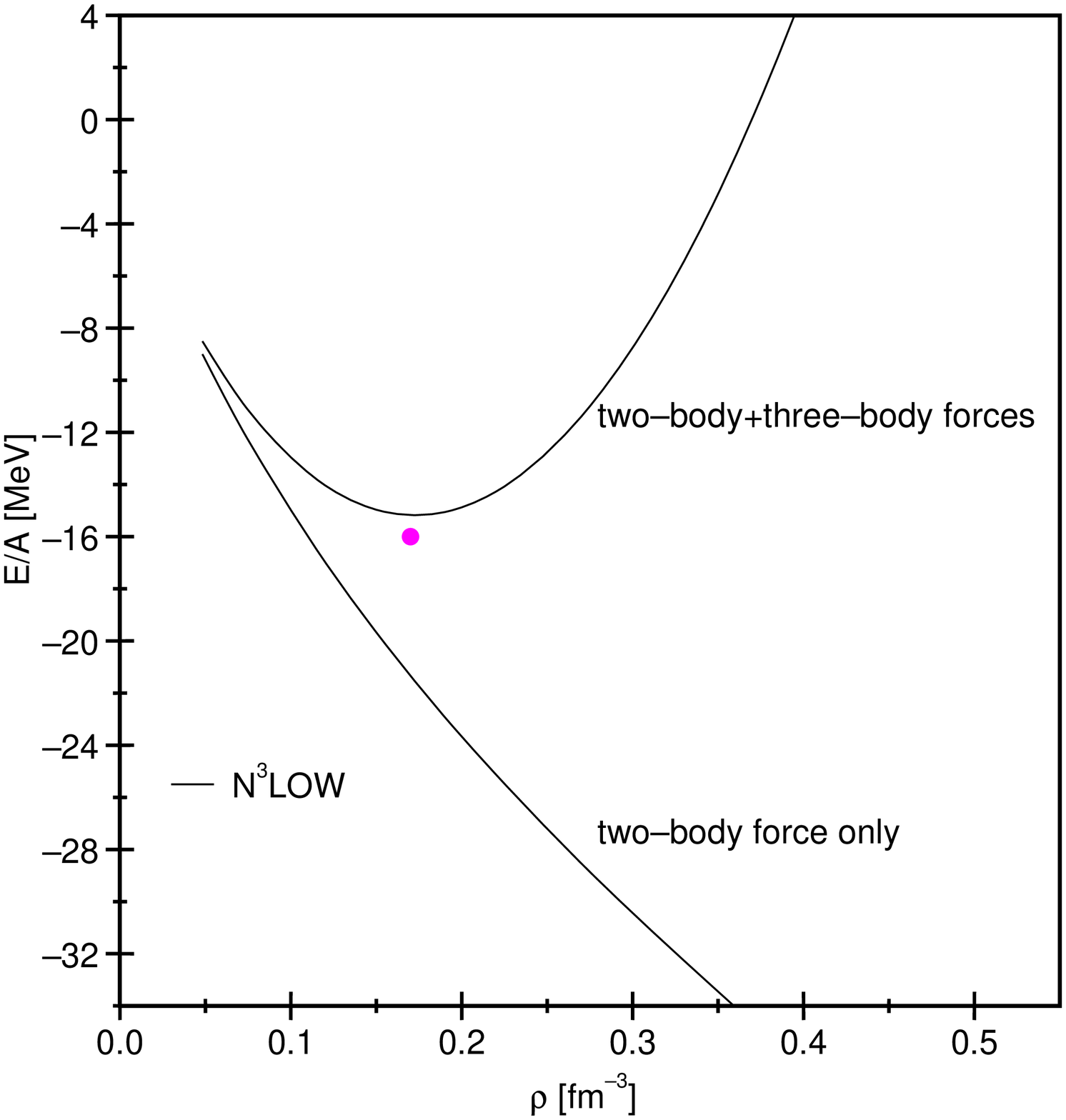} 
\caption{\label{Fig7} (Color online) Same as in Fig. \ref{Fig7}, but for the 
PNM energy per nucleon.}
\end{center}
\end{minipage} 
\end{figure}

In Fig.~\ref{Fig6} and \ref{Fig7} we show, respectively, the
pure-neutron matter (PNM) and symmetric-nuclear matter (SNM) EOS as a
function of density, calculated with and without including  the 3NF
contributions.

From the inspection of the figures, it can be seen that the repulsive 
effect arising from 3NF is consistent in both cases. In particular it can be 
observed that the 3NF contribution to the energy per nucleon in SNM is larger 
than that in PNM, and that its inclusion proves to be crucial to reproduce 
the empirical saturation properties.

\begin{figure}[ht]
\begin{center}
\includegraphics[scale=0.30,angle=0]{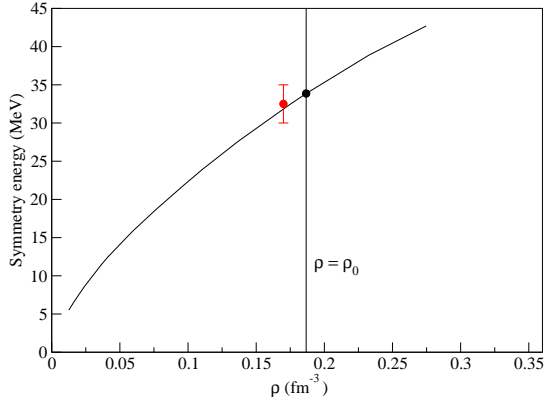}
\caption{(Color online) Calculated symmetry energy as a function of
  density. The red error bar represents the empirical constraint on
  $E_{sym}$ \cite{Dutra2012}. } 
\label{Fig8}
\end{center}
\end{figure}

 The symmetry energy $E_{sym}$ is defined as the strength of the quadratic 
term in an expansion of the energy per particle ($\overline{E}$) in 
asymmetric matter with respect to the asymmetry parameter $\alpha 
= (\rho_n - \rho_p)/(\rho_n + \rho_p)$:
\begin{equation}
\overline{E} (\rho,\alpha) \simeq \overline{E} (\rho,\alpha=0) + 
E_{sym} \alpha^2 + \mathcal{O}(\alpha^4)~~.
\end{equation}

Many microscopic calculations provide a nearly linear behavior 
of $\overline{E} (\rho,\alpha)$ as a function of $\alpha^2$, and this 
justifies the common approximation of defining the symmetry energy as 
the difference between the energy per particle in PNM and SNM, neglecting 
powers beyond $\alpha^2$ in the $\overline{E} (\rho,\alpha)$ expansion.
The density dependence of $E_{sym}$ can be correlated with the neutron skin 
thickness of nuclei and the radius of neutron stars, and 
systematic efforts are ongoing to get better empirical constraints on its 
value, through both laboratory and astrophysical measurements.

In Fig. \ref{Fig8}, our results for the symmetry energy as a function 
of density are reported. 
In this connection, of particular interest is the slope of the
symmetry energy at the saturation density $L$ defined as
\begin{equation}
L = 3 \rho_0 \left ( \frac{\partial E_{sym}}{\partial \rho} \right )_{\rho_0}~~.
\end{equation}
Our calculated value for $L$ is 68 MeV, that has to be compared with the 
not yet firmly constrained empirical value of $L = 70 \pm 25 $MeV 
\cite{Tsang2012}.

\section{Conclusions}\label{conclusions}
In this paper, we have presented nuclear structure calculations, based
on the realistic N$^3$LOW chiral potential, for both finite nuclei and infinite
nuclear matter.

The quality of the results obtained for finite nuclei, and the ability
to provide realistic nuclear matter predictions, makes us confident
that the approach to nuclear structure, based on chiral potentials,
may be a valuable tool to study properties of stable and unstable
nuclear systems.

This excellent agreement paves the way towards an exciting future for
the study of the physics of nuclei.


\begin{thebibliography}{99}
\bibitem{ME11} R. Machleidt and D. R. Entem, {\it Phys. Rep.}, {\bf
    503} (2011) 1.
\bibitem{EHM09} E. Epelbaum,  {\it et al.},
  {\it Rev. Mod. Phys.}, {\bf 81} (2009) 1773.
\bibitem{Wei79} S. Weinberg, {\it Physica}, {\bf 96A} (1979) 327.
\bibitem{EM03} D. R. Entem and R. Machleidt, {\it Phys. Rev. C}, {\bf
    68} (2003) 041001(R).
\bibitem{EGM05} E. Epelbaum,  {\it et al.}, {\it
    Nucl. Phys. A}, {\bf 747} (2005) 362.
\bibitem{Wei92} S. Weinberg, {\it Phys. Lett. B}, {\bf 295} (1992) 114.
\bibitem{Kol94} U. van Kolck, {\it Phys. Rev. C}, {\bf 49} (1994) 2932.
\bibitem{Coraggio07} L. Coraggio,  {\it et al.}, {\it Phys. Rev. C}, {\bf
    75} (2007) 024311.
\bibitem{Bogner02} S. Bogner,  {\it et al.}, {\it Phys. Rev. C}, {\bf
    65} (2002) 051301(R).
\bibitem{Lee80} S. Y. Lee and K. Suzuki, {\it Phys. Lett. B}, {\bf 91}
    (1980) 173.
\bibitem{Marcucci12} L. E. Marcucci,  {\it et al.}, {\it
    Phys. Rev. Lett.}, {\bf108} (2012) 052502.
\bibitem{Coraggio14a} L. Coraggio,  {\it et al.}, {\it Phys. Rev. C}, {\bf 89}
    (2014) 044321.
\bibitem{Coraggio10} L. Coraggio,  {\it et al.},{\it Phys. Rev. C},
    {\bf 81} (2010) 064303.
\bibitem{Coraggio14b} L. Coraggio,  {\it et al.}, {\it Nucl. Phys. A},
    {\bf 928} (2014) 43. 
\bibitem{Coraggio13} L. Coraggio,  {\it et al.}, {\it Phys. Rev. C},
    {\bf 87} (2013) 014322. 
\bibitem{Coraggio12} L. Coraggio,  {\it et al.}, {\it Ann. Phys.}, {\bf
    327} (2012) 2125. 
\bibitem{Kuo90} T. T. S. Kuo and E. Osnes, {\it Lecture Notes in
    Physics}, {\bf 364} (1990).
\bibitem{HKW10} J. W. Holt,  {\it et al.}, {\it
    Phys. Rev. C}, {\bf 81} (2010) 024002.
\bibitem{MacKenzie69} James J. MacKenzie, {\it Phys. Rev.}, {\bf 179}
    (1969) 1002.
\bibitem{ensdf} {\it Data extracted using the NNDC On-line Data
    Service from the ENSDF database}, file revised  as of September
  30, 2015. 
\bibitem{Dutra2012} M. Dutra {\it et al.}, {\it Phys. Rev. C},  {\bf
    85} (2012) 035201.
\bibitem{Tsang2012} M. B. Tsang {\it et al.}, {\it Phys. Rev. C},
  {\bf 86} (2012) 015803. 
\bibitem{Coraggio14c} L. Coraggio, {\it et al.}, {\it Nucl. Phys. A},
    {\bf 928} (2014) 43.
\bibitem{Otsuka10a} T. Otsuka, {\it Phys. Rev. Lett.}, {\bf 105}
    (2010) 032501.
\end{thebibliography}
\end{document}